\documentclass[lettersize,journal]{IEEEtran}
\IEEEoverridecommandlockouts
\usepackage{cite}
\usepackage{amsmath,amssymb,amsfonts}
\usepackage{algorithmic}
\usepackage{graphicx}
\usepackage{booktabs}
\usepackage{tabularx}
\usepackage{textcomp}
\usepackage{xcolor}
\usepackage{comment}
\usepackage{caption}
\usepackage{hyperref}
\usepackage{subcaption}
\def\BibTeX{{\rm B\kern-.05em{\sc i\kern-.025em b}\kern-.08em
    T\kern-.1667em\lower.7ex\hbox{E}\kern-.125emX}}

\newcommand{\orcidauthorA}{\href{https://orcid.org/0009-0008-9900-6125}{\includegraphics[scale=0.05]{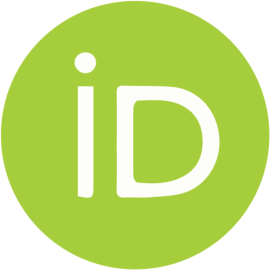}}}
\newcommand{\orcidauthorB}{\href{https://orcid.org/0000-0002-6115-5255}{\includegraphics[scale=0.05]{orcid_16x16.eps}}}
\newcommand{\orcidauthorC}{\href{https://orcid.org/0009-0008-7084-3739}{\includegraphics[scale=0.05]{orcid_16x16.eps}}}
\newcommand{\orcidauthorD}{\href{https://orcid.org/0000-0001-5331-209X}{\includegraphics[scale=0.05]{orcid_16x16.eps}}}
\newcommand{\orcidauthorE}{\href{https://orcid.org/0000-0001-7088-0582}{\includegraphics[scale=0.05]{orcid_16x16.eps}}} 
    
\begin{document}

\title{Co-Existence of Private 5G Network and Wireless Hospital Systems\\

%{\footnotesize \textsuperscript{}Note: Sub-titles are not captured in Xplore and should not be used}
\thanks{This work was partially funded by the Research Council of
Finland via 6GFlagship (grant number 346208) and by the
Connecting Europe Facility (CEF)-funded Hola 5G project
(grant number 101133305).}
}

\author{
    {Mohsin Iqbal Khan$^{\orcidauthorA{}}$, 
    Matti Hämäläinen$^{\orcidauthorB{}}$, 
    Timo J. Mäkelä$^{\orcidauthorC{}}$, 
    Erkki Harjula$^{\orcidauthorD{}}$, 
    and Jani Katisko$^{\orcidauthorE{}}$
    }
    \\
    \thanks{M. I. Khan, M. Hämäläinen, and E. Harjula are with the Centre for Wireless Communications, University of Oulu, Finland (e-mail: \{mohsin.khan, matti.hamalainen, erkki.harjula\}@oulu.fi).}
    \thanks{T. J. Mäkelä and J. Katisko are with Oulu University Hospital, Finland (e-mail: \{timo.j.makela, jani.katisko\}@pohde.fi).}
}

\maketitle

\begin{abstract}
This paper investigates the feasibility of deploying private 5G networks in hospital environments, with a focus on the operating room at the brand new Oulu University Hospital, Finland. The study aims to evaluate the interference risk with other wireless systems, and electromagnetic safety of a private 5G network in the 3.9-4.1 GHz band, while ensuring compatibility with legacy wireless systems, such as LTE and Wi-Fi. We conducted a measurement campaign, employing state-of-the-art instrumentation and a methodology that combined high resolution and long-duration spectrum scans. The results demonstrate no measurable interference between the hospital's private 5G network with adjacent LTE (4G) or Wi-Fi bands, confirming the spectral isolation of the 5G transmissions, and vise versa. Additionally, RF exposure levels in the operating room were found to be well below ICNIRP, WHO, and IEEE safety thresholds, ensuring that the network poses negligible biological risk to patients and hospital staff. The study also proposes spectrum management strategies for private 5G networks in hospitals, focusing on adaptive sensing and guardband planning. These findings provide a solid foundation for the integration of private 5G infrastructure in hospitals environments, supporting digital transformation in patient care without compromising electromagnetic compatibility or patient safety. The results also contribute to ongoing discussions around private 5G network deployments in sensitive sectors and provide actionable guidelines for future hospitals' wireless systems planning.
\end{abstract}


\section{Introduction}
\label{sec:intro}

\begin{figure}[t]
    \centering
    \includegraphics[width=\linewidth]{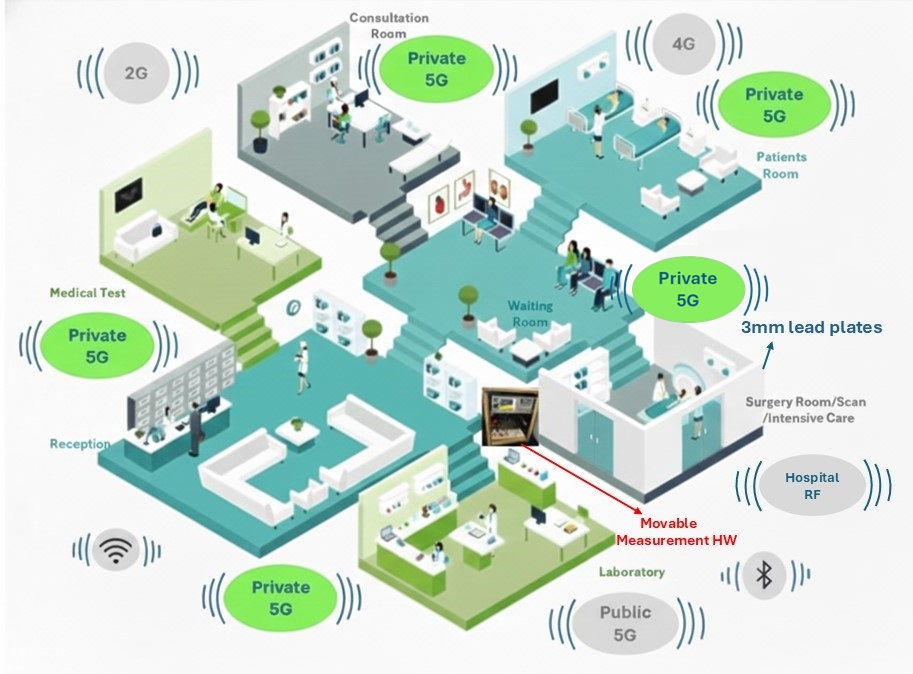}
    \caption{Illustration of the radio frequency landscape in a smart hospital.}
    \label{fig:Smart-hos}
\end{figure}

Wireless communications technologies are increasingly embedded in hospital environments, allowing vital services, such as real-time patient telemetry, wireless infusion pumps, robotic surgery, and high-resolution video transmission in operating rooms~\cite{who_emc, medical_iot_review}. With the emergence of wireless fidelity (Wi-Fi) 6/6E, Bluetooth Low Energy (BLE), and private fifth-generation (5G) networks, hospitals are transitioning toward intelligent, data-driven infrastructures. Fig.~\ref{fig:Smart-hos} (Generated using AI) illustrates the operation room, patient wards, and corridors of a smart hospital, highlighting the deployment of wireless systems such as private 5G, Wi-Fi, and Bluetooth. However, this evolution presents substantial electromagnetic coexistence challenges, especially in environments with life-critical systems that are prone to radio frequency (RF) interference.

The electromagnetic landscape within hospitals is unique: dense, multipath-rich, and filled with legacy and emerging wireless systems that coexist within shared or adjacent spectrum~\cite{coexistence_ieee2021}. Of particular concern is the use of commercial wireless technologies, such as long-term evolution (4G LTE), Wi-Fi, and 5G, in proximity to mission-sensitive devices operating under stringent electromagnetic compatibility (EMC) requirements~\cite{iec_emc}. International guidelines, such as those of the International Electrotechnical Commission (IEC 60601-1-2), the International Commission on Non-Ionizing Radiation Protection (ICNIRP), and International Telecommunication Union (ITU)-T K.91, emphasize the need for careful spectrum planning and on site RF exposure assessment before deploying new technologies in clinical environments~\cite{ICNIRP2020, itu_k91}.

Despite ongoing research in hospital RF monitoring~\cite{rf_meas_hospital, spectrum_clinical_itu}, most existing studies are limited in either spatial granularity, temporal resolution, or technological scope. Few have addressed coexistence of the 0.4-6.1~GHz frequency band in the context of an operational private 5G deployment, particularly under controlled conditions where regulatory compliance, safe exposure, and system-level interference are simultaneously validated. Furthermore, most previous work has not focused on identifying interference-free spectral zones, or ``spectrum islands'', which can be allocated for safe medical data transmission.

This work presents a long-duration spectrum occupancy campaign at the newly constructed Oulu University Hospital  (OYS, Oulun Yliopistollinen Sairaala), where a private 5G network in the 3.9-4.1 GHz band has been deployed for clinical-grade services, such as surgical video transmission and telemetry, as part of the Hola 5G project ~\cite{hola5goulu}. The study quantifies in-band and adjacent-band RF activity using calibrated high-resolution equipment, identifies spectrally isolated bands suitable for medical wireless systems, evaluates the coexistence of private 5G deployment with adjacent LTE and Wi-Fi services, and validates electromagnetic exposure levels against ICNIRP and WHO limits, confirming compliance and safety. Although the electromagnetic safety related to 5G deployments is well understood by RF professionals, we include an assessment of electromagnetic exposure to provide physicians, hospital administrators, and other nontechnical personnel with a clear visualization of the safety of the clinical environment.

\paragraph*{Our Main Contributions} This work presents the first high-resolution 24-hour spectrum occupancy measurement campaign in a new hospital for the 0.4-6.1 GHz band across multiple hospital zones, together with an investigation of spectral interactions between private 5G, LTE, and Wi-Fi services in hospital environments to inform coexistence considerations. In addition, on-site RF exposure measurements of private 5G transmissions in patient-care areas were conducted in accordance with ICNIRP and WHO guidelines, and an occupancy-aware approach was developed to guide frequency coordination and interference management, supporting ultra-reliable, low-latency medical applications.

The remainder of the paper is organized as follows. Section~\ref{sec:related_work} reviews the related literature and existing approaches to the measurements of hospital spectrum and the evaluation of the co-existence. Section~\ref{sec:meas_setup} describes the configuration and deployment of the measurement equipment and the current active wireless systems that were observed in the hospital during the measurement campaign. Section~\ref{sec:methodology} presents the methodology, including measurement protocols, calibration, and data logging. Section~\ref{sec:results} discusses the measurement results, spectral trends, and exposure estimates. Section~\ref{sec:synt_result} discusses the practical and regulatory implications of the results. Finally, Section~\ref{sec:conclusion} concludes the paper.

\section{Related Work}
This section reviews prior research on wireless spectrum monitoring, coexistence, and 5G deployments in hospital environments. It highlights challenges associated with interference, coverage, and electromagnetic exposure, as well as methods proposed to manage spectrum and ensure safe operation.
\label{sec:related_work} 
It highlights the key challenges, including coexistence with legacy wireless systems, coverage limitations, and safe operation, leading into the detailed discussion of RF measurements and spectrum management in the following subsections.
\subsection{Overview of RF Measurements in Hospitals}
Wireless spectrum monitoring in hospital environments presents significant challenges due to the coexistence of diverse medical devices and complex indoor propagation characteristics. Previous studies have primarily focused on characterizing spectrum occupancy and identifying interference sources within limited frequency bands. For instance, Virk et al.~\cite{virk2014spectrum}  performed day-long measurements in the 2.4 GHz industrial, scientific, and medical (ISM) band, revealing high occupancy from overlapping medical and non-medical transmissions, which underscores the need for robust interference management. Similarly, Mucchi et al.~\cite{LorenzoMucchi2014} analyzed spatio-temporal occupancy and modeled aggregate interference, demonstrating persistent utilization in telemetry channels and intermittent activity in others, emphasizing the importance of continuous spectrum awareness. 

This work is partially based on the previous studies \cite{virk2014spectrum}. An 84-day survey by Al Kalaa et al.~\cite{AlKalaa2017} in the 2.4 GHz Wi-Fi band reported low average occupancy (10\%) with sporadic peaks (50\%), indicating that long-duration monitoring is essential to capture transient interference events. More recent work, such as Hossain and Gutierrez~\cite{rf_meas_hospital}, has extended these analyses to broader bands, confirming temporal and spatial variability in hospital spectrum usage.
Recent advancements have incorporated cognitive radio techniques to enhance spectrum efficiency in medical settings. Sodagari et al.~\cite{sodagari2018technologies} explored cognitive radio-enabled wireless body area networks (WBANs), highlighting opportunities for dynamic spectrum access to mitigate interference while maintaining low power consumption. Overall, while prior research has advanced spectrum occupancy characterization and interference detection, it often remains confined to narrow bands or short durations, leaving gaps in comprehensive, wideband assessments suitable for emerging 5G integrations.

\subsection{RF Coexistence and Spectrum Management in Hospitals}
The integration of medical equipment with commercial wireless technologies in hospital settings poses persistent coexistence challenges. The unlicensed 2.4 GHz band, utilized by Wi-Fi and Bluetooth, is particularly susceptible to congestion, potentially interfering with critical devices, such as patient monitors and infusion pumps. Reports from the Medical Records Institute and Mobile Healthcare Alliance~\cite{WirelessPlanning2007} identify radio frequency interference (RFI) as a primary barrier to wireless healthcare adoption, with risks including telemetry disruptions and equipment malfunctions. Case studies have documented interference from nearby cellular devices affecting infusion pumps~\cite{WirelessPlanning2007}. To address these issues, hospitals have implemented spectrum management strategies, including segregation of sensitive communications into dedicated bands. 

In the United States, the Federal Communications Commission (FCC) Wireless Medical Telemetry Service (WMTS) allocates protected bands (e.g., 608-614 MHz and 1.4 GHz) for interference-free patient monitoring~\cite{WMTS_ASHE}, coordinated to minimize external and inter-hospital disruptions~\cite{WirelessPlanning2007}.
However, these dedicated bands offer limited bandwidth and hinder the adoption of advanced networks like private 5G. Alternative approaches suggest shifting devices to less congested 5 GHz Wi-Fi bands; however, emerging unlicensed technologies, such as LTE-U and MuLTEfire, may introduce new contention and interference issues in these shared spectrum environments~\cite{Zinno2018Coexistence}. Literature emphasizes proactive frequency planning and inter-system coordination to mitigate coexistence issues~\cite{WirelessPlanning2007}. Recent surveys on electromagnetic fields (EMF)-aware network planning~\cite{koppel2023survey} highlight the need for regulatory alignment and optimization techniques to balance exposure limits with network performance in dense hospital environments.
\begin{table*}[tb]
\caption{COMPARISON OF KEY STUDIES ON RF SPECTRUM AND COEXISTENCE IN HOSPITAL ENVIRONMENT}
\label{tab:comparison}
\centering
\begin{tabularx}{\textwidth}{XXXX}
\toprule
\textbf{Study} & \textbf{Measurement Focus} & \textbf{Frequency Range} & \textbf{Key Contribution} \\
\midrule
Virk \emph{et al.}~\cite{virk2014spectrum} & Hospital ISM band occupancy & 2.35--2.50 GHz & Long-term occupancy in hospital; no 5G or exposure analysis \\
\midrule
Mucchi \emph{et al.}~\cite{LorenzoMucchi2014} & Hospital ISM band occupancy & 2.35--2.70 GHz & Spatio-temporal occupancy analysis and aggregate interference modelling \\
\midrule
Al Kalaa \emph{et al.}~\cite{AlKalaa2017} & Wireless coexistence testing (Bluetooth/Wi-Fi) & ISM bands & Standards and EMC evaluation for medical device coexistence \\
\midrule
Berlet \emph{et al.}~\cite{berlet2022emergency} & 5G slicing for emergency ultrasound & n/a (mobile/ambulance) & Demonstrated remote diagnostics over 5G; no long-term in-hospital RF validation \\
\midrule
Sodagari \emph{et al.}~\cite{sodagari2018technologies} & Cognitive radio for medical WBANs & Various & Challenges and technologies for spectrum sharing in body area networks \\
\midrule
Abuella \emph{et al.}~\cite{abuella2021hybrid} & Hybrid RF/VLC systems & RF/VLC bands & Survey on topologies, performance, and applications for hybrid networks \\
\midrule
Chiaraviglio \emph{et al.}~\cite{chiaraviglio20225g} & 5G in healthcare from COVID-19 onward & Various & Opportunities and challenges for 5G-enabled healthcare applications \\
\midrule
\textbf{This work} & Private 5G coexistence, exposure safety, spectrum scanning & 0.4--6.1 GHz & 24-hour wideband scan; interference-free 3.9--4.1 GHz; exposure $\ll$ ICNIRP \\
\bottomrule
\end{tabularx}
\end{table*}
\subsection{Private 5G in Hospitals}
Private 5G networks offer transformative potential for hospitals by providing dedicated and secure connectivity for mission-sensitive applications, such as telemedicine, remote surgery, smart healthcare systems, and safe low-latency patient data transmission. Reviews indicate that 5G's high bandwidth and low latency can enhance remote diagnostics and robotic surgery through real-time and high-resolution video streaming~\cite{devi20235g, chiaraviglio20225g}. Pilot projects have demonstrated effective tele-ultrasound over 5G, achieving minimal delays and high-fidelity imaging~\cite{Wu2020}. Berlet et al.~\cite{berlet2022emergency} showcased 5G slicing for emergency ultrasound in mobile settings, though without extensive in-hospital validation.

Despite these advancements, deploying 5G in dense hospital RF environments presents challenges, including interference with sensitive equipment (e.g., magnetic resonance imaging (MRI) machines, infusion pumps) and ensuring robust coverage amid RF obstacles, like metal walls. The Food and Drug Administration (FDA) mandates wireless coexistence analyses for new medical devices to mitigate safety risks in shared bands~\cite{METLabsCoexistence}. Spectrum policy in various countries allocates mid-band frequencies (e.g., 3.7-4.2 GHz) for private 5G, raising concerns about adjacent-band interference~\cite{sorensen2022spectrum}. Empirical studies of private 5G in operational hospitals remain limited, with most relying on simulations or small testbeds that overlook real-world RF complexities~\cite{qureshi2021service}.
\subsection{Electromagnetic Exposure and Safety Standards}
RF exposure is a critical concern in hospitals, where patients and staff are proximate to multiple RF sources, such as  cellular devices, Wi-Fi, and Bluetooth devices. International guidelines from ICNIRP and WHO establish safe limits for RF fields across 100 kHz to 300 GHz, protecting against adverse effects~\cite{ICNIRP2020}. Compliance aims to prevent tissue heating and other substantiated impacts. Ambient RF measurements in hospitals typically show levels far below ICNIRP thresholds; for example, Wi-Fi and cellular signals yield power densities orders of magnitude under 10 mW limits~\cite{Arribas2022}. Bushberg et al.~\cite{Bushberg2025} assessed millimeter-wave exposure from small base stations, confirming compliance with safety standards.
Introducing private 5G requires scrutiny of exposure from higher frequencies and denser infrastructure. 

Few studies quantify exposure in 5G-hospital contexts, as deployments are at an early stage. Prior work focused on lower-power Wi-Fi, Bluetooth, and telemetry systems. Chiaraviglio et al.~\cite{chiaraviglio2021health} analyzed 5G health risks from a communications engineering perspective, concluding no substantiated adverse effects at compliant levels. Dense 5G networks may not increase exposure, as demonstrated by Chiaraviglio et al.~\cite{chiaraviglio2021dense}, due to efficient power distribution. Plets et al.~\cite{plets2024assessment} evaluated uplink exposure from 5G devices, finding levels well within limits. This work contributes empirical data on 5G standalone exposure at 3.99 GHz in a hospital, measuring field strengths during peak traffic against ICNIRP thresholds.
\subsection{Spectrum Isolation and Interference-Free Bands}
Identifying interference-free "spectrum islands" is vital for reliable medical communications. Dedicated allocations, like WMTS bands, provide protected channels for telemetry, minimizing competition with commercial devices~\cite{dedicated_wireless_medical}. Beyond regulated bands, dynamic approaches using cognitive radio enable real-time selection of quiet frequencies, thereby reducing interference risks and enhancing the performance of wireless medical systems~\cite{sodagari2018technologies}. However, achieving isolation in hospitals is challenging due to pervasive RF devices spanning sub-1 GHz to 5 GHz. Private 5G introduces opportunities: if contained without leakage, it serves as a spectrum island for high-priority traffic. Hybrid RF/Visible Light Communication (VLC) systems offer further isolation, leveraging VLC's immunity to RF interference for complementary coverage~\cite{abuella2021hybrid}.
\subsection{Going Beyond the State-of-the-Art}
This work pioneers the evaluation of private 5G networks in a real hospital operational environment, with only a handful of prior projects, such as those in the Franco-German collaboration focusing on time-sensitive networking for tele-services, including robotic surgery and telemedicine~\cite{FrancoGerman5GEcosystem}, and in Hong Kong for smart hospital applications enabling remote consultations, training, and telemedicine via low-latency 5G technologies~\cite{GSMA2022}, conducting experimental 5G work. The novelty of our work lies in long-duration (24-hour), high-resolution spectrum measurements across 0.4--6.1 GHz, assessing coexistence between private 5G (3.9--4.1 GHz) and adjacent Wi-Fi/LTE services, and empirically validating RF exposure against ICNIRP/WHO thresholds. Table~\ref{tab:comparison} summarizes key prior studies on hospital spectrum usage, coexistence, and interference management, contrasting them with this work.

\begin{figure*}
    \centering
    \includegraphics[width=1\linewidth]{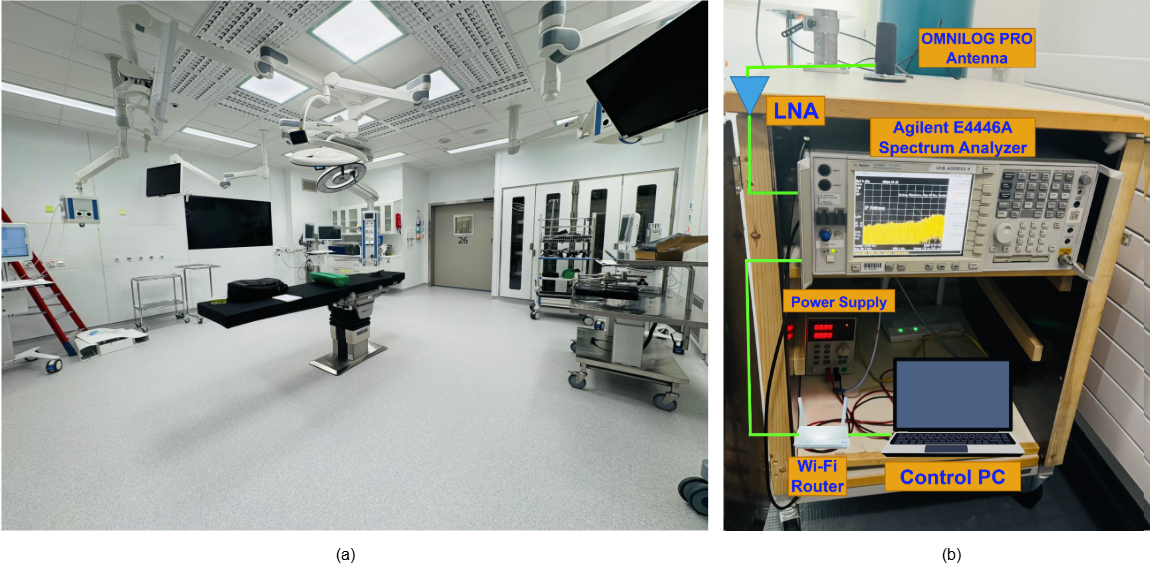}
    \caption{(a) Operation room at the New Hospital facility; (b) Measurement setup architecture used for the spectrum campaign.}
    \label{fig:Oper_Meas}
\end{figure*}

\section{Measurement Setup}
\label{sec:meas_setup}

This section represents the architecture of the RF measurement system utilized for capturing spectrum occupancy within the operating room in OYS, as illustrated in Fig.~\ref{fig:Oper_Meas}a. The system was engineered to enable high-resolution, interference-resistant signal acquisition across the frequency band from 400~MHz to 6.1~GHz, adhering to regulatory standards and accommodating the demands of medical and research applications. Precision instrumentation, optimized RF front-end elements, and automated control mechanisms were integrated to support repeatable, high-fidelity data acquisition, capable of monitoring both persistent and transient RF phenomena in a dynamic hospital setting. The overall measurement configuration is shown in Fig.~\ref{fig:Oper_Meas}b.\\

\noindent\textbf{Current Systems/Networks in Hospital:} During~our~measurement campaign in the hospital, we observed multiple active wireless systems spanning the frequency range from 400 MHz to 6.1 GHz. The identified bands included legacy GSM services in the 760-960 MHz range, LTE and new radio (NR) carriers in 1.7-2.6 GHz (Bands 1, 3, and 7), unlicensed ISM operations (Wi-Fi and Bluetooth) around 2.3-2.4 GHz, and Wi-Fi 5/6 activity within 5.1-5.7 GHz. In addition to these commercial and unlicensed deployments, the hospital also operates a pre-installed private 5G network in the 3.9-4.1 GHz band, which was dedicated to medical broadband applications and experimental services. It should also be noted that 3G networks were shut down in Finland by the end of 2024.\\

\noindent\textbf{Measurement Instrumentation and Frequency Coverage:} The core instrument employed was the Agilent E4446A spectrum analyzer (SA)~\cite{keysightE4446A}, a heterodyne receiver capable of wideband signal acquisition from 3~Hz to 44~GHz. In this study, the SA was configured to span the 400~MHz to 6.1~GHz range, encompassing critical wireless bands prevalent in hospital environments, including medical telemetry (400--608~MHz), ISM bands (902--928~MHz, 2.4--2.5~GHz, and 5 GHz), private 5G pilot allocations (3.9--4.1~GHz), Wi-Fi 5 (5.725--5.875~GHz) and Wi-Fi 6 (5.925--6.1~GHz).

Network integration of the SA was achieved by configuring its IP settings via the system interface. The System Settings menu facilitated the establishment of fundamental network parameters, followed by selection of the Config I/O option to activate Ethernet communication. A static IP address was manually assigned to align with the control laptop's network configuration. The SA operated in logarithmic power detection mode to yield power spectral density (PSD) profiles. External signal amplification was provided by a cascaded low-noise amplifier (LNA) chain, essential for augmenting system sensitivity while maintaining linearity across the signal path.\\

\noindent\textbf{RF Frontend and Antenna System:} RF signals were received via the OmniLOG PRO antenna, offering omnidirectional coverage from 150~MHz to 18~GHz. This antenna was chosen for its spectral flatness, high sensitivity, and robustness in multipath-rich indoor environments characteristic of hospitals. It exhibits vertical linear polarization and a voltage standing wave ratio (VSWR) below 2.5:1, ensuring effective impedance matching over the operational frequencies.~\cite{aaronia_omnilogpro}. The antenna-to-analyzer interconnection employed SubMiniature version A (SMA) coaxial cables, renowned for their application in high-frequency and microwave systems due to threaded coupling and 50 $\Omega$ impedance.~\cite{moco2024_rf_sma}\\

\noindent\textbf{Automation and Remote Control:} Remote operation and automated measurements were enabled through a MATLAB-based control suite interfacing with the SA over an Ethernet local area network (LAN). This was supported by a TeleWell TW-EAV510AC router configured exclusively in wired mode to mitigate electromagnetic interference (EMI) from wireless transmissions. Static IP addressing ensured robust connectivity between the control laptop and the SA, thereby stabilizing the measurement framework~\cite{telewell_products}.

Measurement configurations were scripted in MATLAB using the Standard Commands for Programmable Instruments (SCPI) protocol. Separate scripts managed parameter setup and the execution of sweeps, data retrieval, and storage. Data integrity was upheld through pre-session calibration of the SA using its internal routines, promoting measurement traceability and accuracy. This automation paradigm enhanced repeatability and reduced variability attributable to human intervention.

\section{Methodology}
\label{sec:methodology}

This section presents the methodological framework for evaluating the coexistence of wireless devices and electromagnetic safety in hospital settings. The design prioritizes accuracy and reproducibility, conforming to established practices while mitigating challenges inherent to clinical environments. This approach facilitates the integration of technologies, such as private 5G without jeopardizing patient safety or disrupting incumbent medical systems. It aligns with IEC 60601-1-2~\cite{iec_emc}, ITU-T K.91~\cite{itu_k91}, and ICNIRP guidelines~\cite{ICNIRP2020}, extending methodologies from prior RF studies in hospitals~\cite{rf_meas_hospital, coexistence_ieee2021}.

\subsection{Measurement Strategy and Site Selection}

Measurements were conducted in the operating room during ongoing surgery, selected for its importance in requiring interference-free communications—particularly for real-time telemetry and surgical video streaming. The environment included typical surgical equipment, such as patient monitors, infusion pumps, and flex arms. This locale serves as an exemplary venue for scrutinizing the implications of wireless technologies like private 5G amid electromagnetically sensitive equipment. The frequency coverage extended from 400~MHz to 6.1~GHz, incorporating bands for medical telemetry, ISM applications, Wi-Fi, commercial mobile networks,  and private 5G trials (3.9--4.1~GHz). Sessions spanned at least 24 hours to encapsulate steady-state and transient RF dynamics, accounting for diurnal variations in clinical activities and spectrum utilization.

\subsection{Controlled Instrument Configuration and Repeatability}
Consistency across all measurement sessions was maintained through a validated MATLAB-based automation framework that controlled the entire instrumentation chain. Before each 24-hour acquisition, the spectrum analyzer and peripheral equipment were initialized with standardized configuration parameters summarized in Table~\ref{tab:meas_params}. These settings—including fixed frequency span, resolution bandwidth, and sweep timing—were pre-validated to ensure an optimal balance between spectral resolution and acquisition efficiency, enabling the reliable detection of low-power emissions while managing extensive data volumes.

The automated measurement scripts minimized operator intervention by executing parameter configuration, sweep control, data retrieval, and logging in a single integrated workflow. This approach reduced human-induced variability that often affects long-duration measurements. Calibration preceded every acquisition cycle using the instrument’s internal reference routine, ensuring traceability of the recorded power levels across all sessions.

\subsection{Environment Control and External Interference Mitigation}

To minimize measurement artifacts and ensure that the recorded signals accurately represented in-hospital wireless activity, strict environmental controls were implemented throughout the measurement campaign. On the measurement laptop, all wireless interfaces, including Wi-Fi and Bluetooth, were disabled prior to data collection to prevent unintended emissions from the control system itself. Medical personnel were informed in advance and instructed to avoid introducing unnecessary personal wireless devices into the measurement zones during the sessions.

The control network connecting the spectrum analyzer to the MATLAB automation framework operated exclusively over wired Ethernet links, eliminating any potential RF emissions from wireless connectivity. Following measurement initialization, the operator remained outside the measurement area to ensure that no unintended interference was introduced during data acquisition. The combined use of personnel awareness, wired communication links, and controlled scheduling significantly enhanced signal fidelity, measurement repeatability, and the overall reliability of the resulting dataset.

\begin{table}[!t]
\centering
\caption{KEY MEASUREMENT PARAMETERS}
\label{tab:meas_params}
\begin{tabular}{l c}
\hline
\textbf{Parameter} & \textbf{Value/Description} \\
\hline
Frequency Range (MHz) & 400 - 6100 \\
Resolution Bandwidth (RBW) & 1 \, MHz \\
Number of Frequency Points per Sweep & 5701 \\
Sweep Time & $\approx\,9$ ms \\
Number of Sweeps per Measurement Instance & 1000 \\
Frequency Resolution & 999.82 kHz \\
Measurement Duration per Session & $\approx\,24$ hours \\
Data Acquisition per 24-h Measurement Session & $\approx\,645$ MB \\
\hline
\end{tabular}
\end{table}

\section{Results}
\label{sec:results}
This section presents the empirical findings from the measurement campaign conducted in the operating room of OYS, along with a detailed analysis of the implications of these findings. The results focus on the spectral isolation of the private 5G network and the assessment of electromagnetic safety, providing a clear understanding of its performance and compliance with relevant safety standards.

\begin{figure}[!b]
    \centering
    \includegraphics[width=1\linewidth]{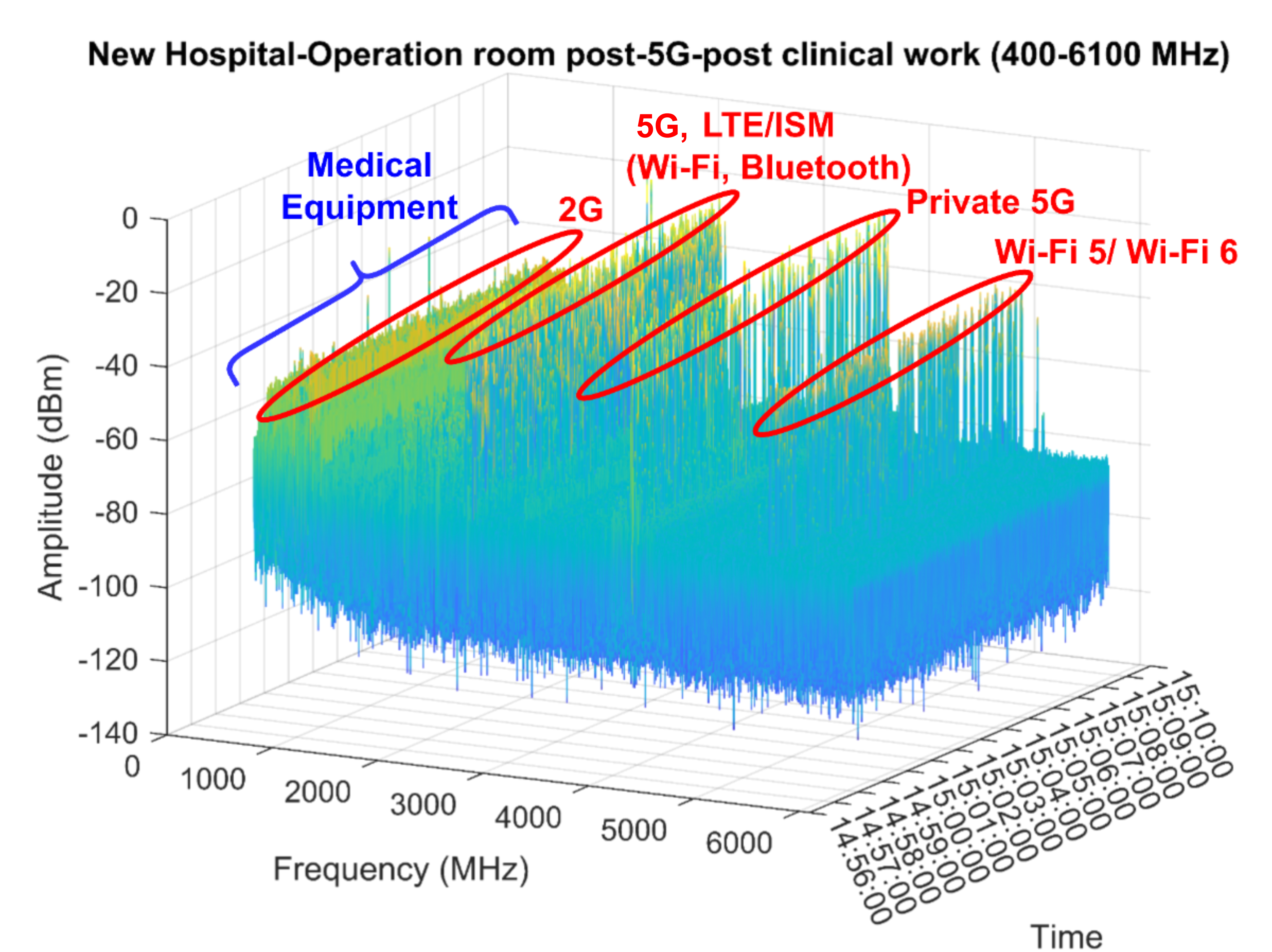}
    \caption{Illustration of 5G network isolation in the operation room.}
    \label{fig:isolation}
\end{figure}
\begin{figure}[!b]
    \centering
    \includegraphics[width=1\linewidth]{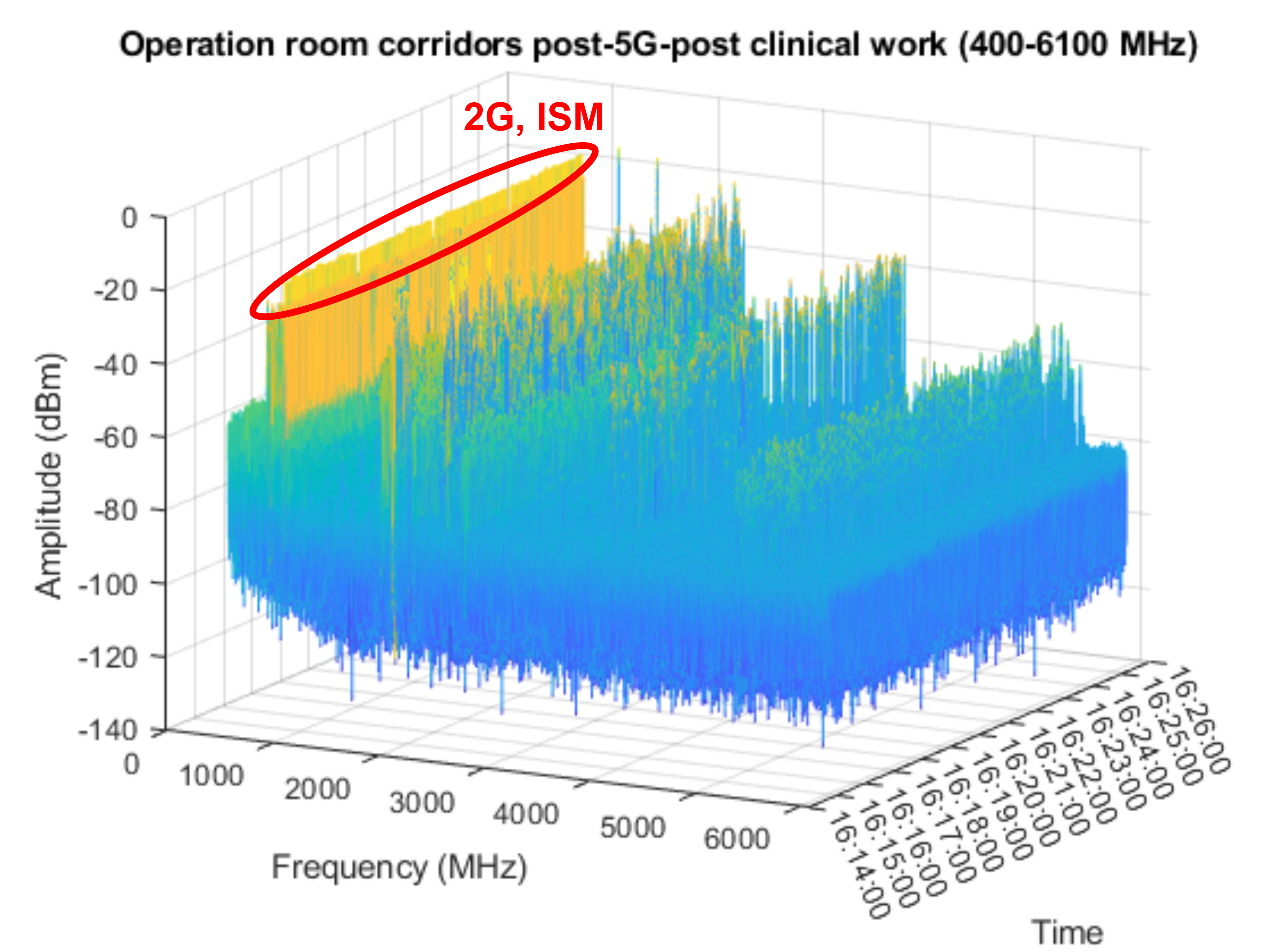}
    \caption{Received power spectral density in the new hospital corridors outside operation room.}
    \label{fig:NewHos_corridor}
\end{figure}
\subsection{Spectral Isolation and Reliable Coexistence}
\label{sec:spect_iso}
% Most of the medical devices operates around 900MHz, while 5G is around 3.9GHz--4.1GHz, so no chance of mutual interference, still the isolation with the adjucent ISM bands is assest below ...

The key finding from this study is the spectral isolation achieved by the 3.9-4.1 GHz band used for the private 5G network. Throughout the 24-hour measurement period, no detectable interference was observed from the 5G transmissions in the adjacent LTE Band 7 (2.6 GHz) or Wi-Fi bands (2.4 and 5 GHz). Specifically, the power levels at 2.574 GHz (LTE B7 uplink) and 5.2 GHz (Wi-Fi 5/6) remained at ambient noise levels during active 5G transmission. This outcome aligns with expectations given that the 5G band is separated by over 1 GHz from LTE Band 7 and approximately 800 MHz from the 5.2 GHz Wi-Fi band. These findings confirm that out-of-band emissions from the 5G network are well-contained, a characteristic that is vital for the safe coexistence of these technologies in sensitive environments, like hospitals.

Although most of the medical devices and instruments typically operate around or below 900 MHz and the private 5G operates in 3.9 GHz-4.1 GHz band, there is no possibility of mutual interference. Still, the spectral isolation of private 5G network with adjacent bands is assessed. Fig.~\ref{fig:isolation} confirms the 5G transmissions (3.9-4.1 GHz) do not leak into adjacent LTE or Wi-Fi bands, supporting the spectral isolation of the 5G band. The measured isolation is consistent with 3rd Generation Partnership Project (3GPP)/European Telecommunications Standards Institute (ETSI) standards, which stipulate that 5G base stations must adhere to steep out-of-band emission masks, ensuring minimal unwanted emissions outside their designated band.

Furthermore, the study underscores the importance of maintaining wide guard bands between mission-sensitive systems. The 1.3-1.4 GHz gap between the 5G band and LTE Band 7, and the 800 MHz gap to Wi-Fi, exceed typical duplex guard-band requirements. These wide gaps contribute to ensuring that the 5G network operates as an isolated spectrum island relative to the neighboring systems, thus facilitating safe operation without disrupting vital and sensitive medical equipment. From a clinical perspective, the results highlight the feasibility of using a private 5G network in the hospital for real-time applications in operating rooms and patient wards. The findings suggest that well-engineered 5G systems, particularly those with high-frequency assignments, can coexist harmoniously with existing technologies in hospital environments, ensuring both operational integrity and patient safety.
\subsection{Physical Isolation of the Operating Room Environment}
\label{sec:physical_isolation}
% ---- Paragraph 1 starts ----
While Section~\ref{sec:spect_iso} demonstrated spectral isolation of the private 5G band from adjacent frequency services, it is equally important to assess the physical isolation provided by the operating room structure itself. Modern operating rooms in the hospital under study were designed with electromagnetic shielding materials, such as 3 mm lead plates within walls, to reduce electromagnetic interference from external wireless systems. To quantify the shielding effectiveness, additional measurements were conducted in the hospital corridors adjacent to the operating room using the same methodology described in Section~\ref{sec:meas_setup}. 
% ---- Paragraph 1 ends ----

% ---- Paragraph 2 starts ----
As shown in Fig.~\ref{fig:NewHos_corridor}, sub-1 GHz (2G and ISM) band at around 800--900 MHz exhibits significantly higher received power levels and occupancy in the corridors compared to those measured inside the operating room, as observed in Fig.~\ref{fig:isolation}. Consequently, the combination of the spectral isolation of the private 5G band and the structural shielding of the operating room provides a robust interference-free environment for mission-sensitive medical applications, such as real-time imaging, robotic-assisted surgery, and low-latency data transfer.
% ---- Paragraph 2 ends ----

\subsection{Electromagnetic Safety and RF Exposure}
% ---- Paragraph 1 starts ----
The electromagnetic safety of the private 5G network in the operating room was thoroughly assessed. RF exposure levels in the 3.9--4.1 GHz band were found to be well below internationally recognized safety thresholds. The maximum received power of -16 dBm corresponds to a power density of approximately 0.05–0.1 mW, which is far below the limits set by ICNIRP and WHO for general public exposure~\cite{itu_emc, who_emc}. Even under worst-case assumptions (isotropic reception), exposure remains within safe limits, confirming the negligible biological risk posed by the in-hospital deployment of 5G. These results align with previous studies, indicating that properly engineered 5G systems do not pose significant health risks in hospital settings~\cite{WHO2020, ICNIRP2020}.
% ---- Paragraph 1 ends ----
\begin{figure}
    \centering
    \includegraphics[width=1\linewidth]{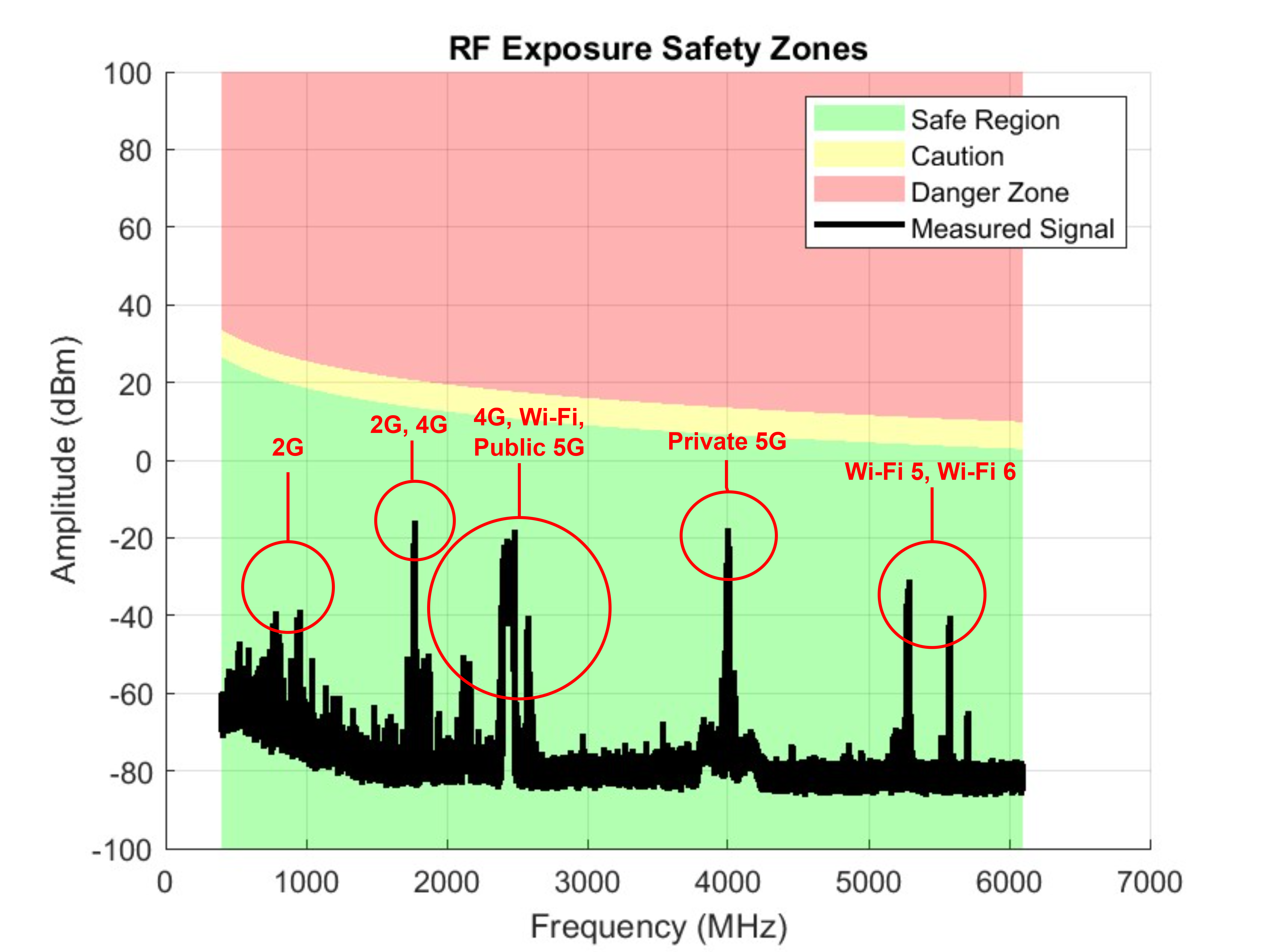}
    \caption{Measured maximum signal amplitudes compared with the safety standards of ICNIRP, WHO, ITU, and IEEE.}
    \label{fig:safety}
\end{figure}
% ---- Paragraph 2 starts ----

Furthermore, the assessment included both temporal and spatial variations in the operating room, ensuring that short-term peaks as well as cumulative exposure over the 24-hour measurement period remained within safe boundaries. This comprehensive evaluation is particularly important in hospital environments, where patients and staff may spend extended periods near active 5G transmitters. Fig.~\ref{fig:safety} illustrates that the peak received power remains comfortably within the green zone of RF exposure safety limits, significantly lower than the typical general public safety limit of 10 mW for frequencies above 400 MHz~\cite{ICNIRP2020}. The results provide strong evidence that the private 5G network operates safely under real-world conditions, ensuring both patient and staff safety during clinical operations.
% ---- Paragraph 2 ends ----

% ---- Paragraph 3 starts ----
In addition to the measured RF exposure levels, the physical placement of the wireless equipment further reduces potential risks. The Wi-Fi and private 5G radios in the operating rooms were mounted on the ceiling, approximately 5 meters above the floor. This configuration ensures that patients and staff remain at least 1.5 meters away from the radios at all times, minimizing direct exposure while maintaining reliable wireless coverage throughout the room.
% ---- Paragraph 3 ends ----

\section{SYNTHESIS OF THE RESULTS}
\label{sec:synt_result}
This section consolidates the key findings from the measurement campaign and frames them in terms of practical, regulatory, and policy implications. It highlights how the empirical results inform spectrum management strategies, safe network deployment, and the integration of private 5G networks in hospital environments.
\subsection{Spectrum Management and Practical Implications}

From a practical standpoint, the results of this study provide valuable insights into spectrum management strategies for private 5G networks in hospitals. Adaptive sensing and thresholding techniques can be employed to dynamically adjust channel sensing based on real-time spectrum conditions. For example, the system can be adjusted to focus on stronger signals while filtering out background noise, allowing hospitals to minimize false detections and improve the reliability of spectrum monitoring, as suggested in previous studies.~\cite{al2016characterizing}. This approach ensures that real-time spectrum monitoring can be both efficient and reliable, even in a dynamic hospital environment with fluctuating RF conditions.

Additionally, the study reinforces the importance of guard bands and planned separation between vital 5G channels and known high-activity bands. For example, the 5G band in the hospital was carefully separated from the 2.4/5 GHz ISM bands, which are heavily used by Wi-Fi and Bluetooth devices. These efforts are consistent with regulatory guidelines that require careful frequency planning and interference mitigation, ensuring that 5G transmissions do not spill into adjacent medical telemetry or Wi-Fi channels~\cite{ITU2025}. Moreover, this study supports the recommendation to avoid using ISM bands for mission-sensitive 5G services, as these bands are heavily congested and unpredictable, especially in hospital settings where the number of wirelessly connected devices is growing rapidly.

The methodology outlined in this study, incorporating long-duration spectrum scans and adaptive thresholding, provides a reproducible model for spectrum management in sensitive environments. Hospitals can use these guidelines to make informed decisions about frequency allocation, interference avoidance, and safe device integration. These insights are particularly valuable as the demand for medical IoT, real-time diagnostics, and wireless telemetry continues to grow, and the need for reliable, interference-free wireless systems becomes more important.

\subsection{Clinical Implications}

Although 5G technology has been declared safe by international health authorities, concerns persist among healthcare professionals about its potential impact on patients, staff, and the hospital environment. These concerns often arise from uncertainties about electromagnetic exposure, the coexistence of multiple wireless systems, and the potential for unexpected interference with existing equipment. Addressing these concerns requires not only the adherence to international safety standards, but also empirical evidence from real-world hospital settings, which can provide practical reassurance to clinical personnel and administrators.

The results confirm that the deployed private 5G network operates safely and does not introduce measurable interference with existing hospital wireless systems. This gives healthcare professionals and administrators confidence that modern wireless technologies can be integrated without compromising patient safety or daily operations. Importantly, the electromagnetic safety assessment shows that the levels of RF exposure are well below internationally recognized thresholds, ensuring negligible biological risk to patients and staff. This reinforces confidence in the safe deployment of 5G in clinical environments.

More broadly, these findings provide clear and evidence-based information on the safety and interference of 5G in a sensitive healthcare environment. The study helps address common misconceptions and public concerns about 5G technology. This understanding encourages informed discussions and greater acceptance of advanced wireless networks in critical infrastructure, showing that the potential risks often highlighted in public debates are largely controlled when proper planning and regulatory standards are followed. Overall, the study demonstrates how technical measurements can be translated into practical guidance for healthcare professionals. It shows that empirical evidence can support safe adoption, shape policy decisions, and build confidence among clinicians and the public alike, confirming that private 5G networks can be integrated into hospital environments without compromising safety, reliability, or day-to-day clinical operations.

\subsection{Regulatory and Policy Implications}
The results of this study have broader regulatory and policy implications, particularly in relation to the adoption of dedicated spectrum for private networks in industries, as they demonstrate that private 5G networks can operate safely and effectively in hospital environments. This study aligns with ongoing international efforts to promote private network adoption in sectors such as hospitals, aviation, and manufacturing~\cite{itu_k91}. By demonstrating the feasibility of 5G coexistence, this work contributes to shaping future spectrum policies that support innovative, secure, and safe 5G network deployments across vital strategic infrastructures. The findings also underscore the potential for spectrum policy reforms that could facilitate the deployment of 5G networks in hospitals without disrupting existing systems. This supports the broader push for shared licensed spectrum or local spectrum regimes that allow private network rollouts in essential sectors.

This study provides empirical evidence that private 5G networks can be deployed safely and effectively in hospital environments, causing minimal interference while maintaining compliance with international safety standards. The findings validate that a well-engineered private 5G network can coexist harmoniously with legacy wireless systems, making it feasible to use for mission-sensitive healthcare applications. The results further contribute to the development of actionable spectrum management guidelines, which can help hospitals to deploy private 5G infrastructure without disrupting existing systems. Moreover, the demonstrated electromagnetic safety of the private 5G network in the hospital environment supports the ongoing digital transformation in healthcare, paving the way for more innovative applications and improved patient care outcomes.

\section{Conclusion}
\label{sec:conclusion}
This study evaluated the feasibility and safety of deploying a private 5G network in the operating room of Oulu University Hospital. A high-resolution, 24-hour measurement campaign was conducted across the 0.4–6.1 GHz spectrum to assess spectral occupancy, coexistence, and electromagnetic exposure. The analysis focused on interactions between the private 5G network in the 3.9–4.1 GHz band and existing LTE and Wi-Fi systems. The results confirm that the private 5G network remains spectrally isolated, with no measurable interference observed in adjacent LTE or Wi-Fi channels, validating compliance with 3GPP/ETSI emission requirements. Furthermore, the results also show that there are no coexistence issues in other frequency bands beyond 5G’s, confirming stable operation across the measured spectrum.

Electromagnetic exposure measurements indicate that the maximum received power of -16 dBm corresponds to power densities far below internationally recognized safety thresholds, demonstrating compliance with ICNIRP, WHO, and IEEE guidelines. These findings provide empirical evidence that private 5G networks can operate safely in sensitive hospital environments, ensuring that patient care and critical medical equipment remain unaffected. Overall, the study demonstrates that well-engineered private 5G deployments can coexist harmoniously with legacy wireless systems, maintain spectral isolation, and operate within safe exposure limits, providing a solid foundation for the integration of private 5G infrastructure in hospital environments. Finally, while this study focused on methodology and aggregate results, detailed analysis of channel occupancy patterns can be extracted from the collected data and will be addressed in our future work.

%\section{Acknowledgments}
%\label{sec:Ack}
% This work was partially funded by the Research Council of Finland via 6GFlagship (grant number 346208) and by the Connecting Europe Facility (CEF)-funded Hola 5G project (grant number 101133305).
\clearpage
\bibliographystyle{IEEEtran}
\bibliography{readme.bib}
\end{document}